# Solubility of ionic surfactants below their Krafft point in mixed micellar solutions: Phase diagrams for methyl ester sulfonates and nonionic cosurfactants

Krassimir D. Danov[*,1], Rumyana D. Stanimirova[1], Peter A. Kralchevsky[1],

Tatiana G. Slavova[1], Veronika I. Yavrukova[1], Yee Wei Ung[2], Emily Tan[2], Hui Xu[2],

Jordan T. Petkov[1,3]

[1] *Department of Chemical & Pharmaceutical Engineering, Faculty of Chemistry & Pharmacy, Sofia University, 1164 Sofia, Bulgaria*

[2] *KLK OLEO, KL-Kepong Oleomas Sdn Bhd, Menara KLK, Jalan PJU 7/6, Mutiara Damansara, 47810 Petaling Jaya, Selangor Dalur Ehsan, Malaysia*

[3] *Biological Physics, School of Physics and Astronomy, The University of Manchester, Schuster Building, Oxford Road, M13 9PL, UK*

———————

\* Corresponding author

  *E-mail address*: kd@lcpe.uni-sofia.bg (K.D. Danov).

ABSTRACT

*Hypothesis:* Many ionic surfactants with wide applications in personal-care and house-hold detergency show limited water solubility at lower temperatures (Krafft point). This drawback can be overcome by using mixed solutions, where the ionic surfactant is incorporated in mixed micelles with another surfactant, which is soluble at lower temperatures.

*Experiments:* The solubility and electrolytic conductivity for a binary surfactant mixture of anionic methyl ester sulfonates (MES) with nonionic alkyl polyglucoside and alkyl polyoxyethylene ether at 5 °C during long-term storage were measured. Phase diagrams were established; a general theoretical model for their explanation was developed and checked experimentally.

*Findings:* The binary and ternary phase diagrams for studied surfactant mixtures include phase domains: mixed micelles; micelles + crystallites; crystallites, and molecular solution. The proposed general methodology, which utilizes the equations of molecular thermodynamics at minimum number of experimental measurements, is convenient for construction of such phase diagrams. The results could increase the range of applicability of MES–surfactants with relatively high Krafft temperature, but with various useful properties such as excellent biodegradability and skin compatibility; stability in hard water; good wetting and cleaning performance.

## 1. Introduction

The sulfonated methyl esters (MES) are produced from renewable palm-oil based materials [1–4]. The commercial MES surfactants have typically even alkyl chain lengths, from $C_{12}$ to $C_{18}$, and they are denoted below C$n$-MES. MES have excellent biodegradability and biocompatibility [5] and have been promoted as alternatives to the petroleum-based surfactants [6–9]. The sulfonated methyl esters have attractive properties for various applications [10–16]: very good wetting and detergent powers with low viscosity of their aqueous solutions; suitable for cleaning formulations, such as phosphate-free detergent powders; excellent water hardness stability allows them to be used in hard-water regions; very good ability to dissolve calcium-soap scum; excellent skin compatibility that makes them potentially very good for hand dishwashing formulations and body care products; viscous formulations, in mixture with nanoparticles. The recent studies on the rheological behavior of mixed MES and betaine solutions [17] and on the oil drop deposition on solid surfaces [18] have shown a possible wide application for shampoo systems.

The interfacial properties of C$n$-MES and their critical micelle concentrations (CMC) have been studied by the surface tension, electrolytic conductivity, and neutron reflectivity measurements [9,19–24]. The experimental data and theoretical interpretations show that C$n$-MES molecules exhibit typical behavior for ionic surfactants: CMC decreases with the increase of number of carbon atoms in the alkyl chain, $n$, and with the salt concentration; the adsorption at the CMC is 3.4 μmol/m$^2$ and the excluded area per molecule is 37 Å$^2$; the binding energy of $Ca^{2+}$ ions to the headgroup of MES is comparable to that of $Na^+$ ions and considerably smaller than to the headgroups of linear alkylbenzene sulfonates. One of the possible impurities in commercial MES is the fatty acid sulfonate (disalt), which also forms at large enough values of pH [2,25,26]. For surfactant concentrations below 200 mM and low concentrations of added salt, the micelles are spherical with aggregation numbers from 57 (for C12-MES) to 90 (for C16-MES) [27]. The C$n$-MES forms wormlike micelles with the rise of added salt concentration and in mixtures with betaine [17,27,28].

The increase of the length of MES alkyl chain leads to the increase of the Krafft temperature, $T_K$ [29]: $T_K$ = 28 $^o$C for C16-MES; $T_K$ = 41 $^o$C for C18-MES. For that reason in many cases, the eutectic mixtures of C16- and C18-MES are used [24]: for example the Krafft temperature of C16/C18 (3/1) weight to weight fractions decreases to 15 $^o$C. For temperatures lower than 15 $^o$C, all kind of C16-C18-MES mixed solutions are turbid at large enough concentration because of the formation of MES precipitates (MES-crystals) instead of micelles. An efficient way to increase the solubility of long chain length MES is to incorporate the MES molecules in the micelles of other surfactants, which do not precipitate at low temperatures. It is shown in the literature that the solubility of fatty acids and alcohols increases considerably in anionic (SLES) and zwitterionic (CAPB) micellar surfactant solutions [30,31].

Our goal in the present study is to obtain experimentally the solubility limits of individual C$n$-MES components and their mixtures in pure water at low temperature of 5 $^{o}$C (Section 3). The increase of the solubility of the C16-MES and C18-MES and their mixtures in micellar commercial (Pareth-7 and Glucopon) surfactant solutions is detected by the measurements of the saturation mole fractions of MES in respective micelles (Section 3). The obtained experimental data allow calculating the complete set of physicochemical constants of MES and nonionic surfactants in the bulk and micelles. The theoretical approach from Refs. [30,31] is applied to calculate the phase diagrams for individual MES components in micellar surfactant solutions (Section 4). The four phase diagram domains are separated by the four phase separation lines, which intersect in one quadruple point. The theory is generalized in Section 5 for mixtures of two partially soluble components and one nonionic surfactant. The number of domains in the essential 3D phase diagrams increases by 2 new regions containing precipitates from the both MES. As a result, quantitative descriptions of all seven phase separation lines and two quadruple points are achieved. The results could be of interest for any application, in which mixed micellar solutions of nonionic surfactants and MES mixtures are used, and the solutions should be clear at low temperatures.

## 2. Materials and methods

The following sulfonated methyl esters (C$n$-MES), produced by KLK OLEO, were used (Fig. 1): myristic (C14-MES) with molecular weight $M_w$ = 344 g/mol and the critical micelle concentration (CMC) 3.68 mM [21] at 25 $^{o}$C; palmitic (C16-MES), $M_w$ = 372 g/mol, and CMC = 1.02 mM [21] at 25 $^{o}$C; mixtures of palmitic and stearic (C18-MES, $M_w$ = 400 g/mol) with C16-MES/C18-MES weight to weight fractions 80/20 and 60/40 and CMC < 1 mM at room temperature. In all experiments the pH of MES solutions was adjusted to 5.5. The used C14-MES and C16-MES samples have been characterized by liquid chromatography-mass spectrometry (LC/MS) analysis. The purity of C14-MES is 97.9 %, of C16-MES is 99.1% [21], and the active substances of C16-MES/C18-MES mixtures are > 92%.

The ability of two nonionic surfactants to increase the MES solubility at low temperature of 5 $^{o}$C was studied (Fig. 1). Pareth-7 (Imbentin-AG/124S/070, product of KLK

OLEO) is a nonionic surfactant with a linear alkyl chain of 12-14 carbon atoms, a hydrophilic head of 7 ethoxy groups, average molecular weight 516 g/mol, natural pH = 6.2, > 99% active. Glucopon 225 DK, product of BASF, has average molecular mass 390 g/mol, natural pH = 6.8, 8-10 carbon atoms in the linear alkyl chain, 1.2-1.5 glucoside groups in the hydrophilic part, 69.5% active. All chemicals were used as received, without additional purification.

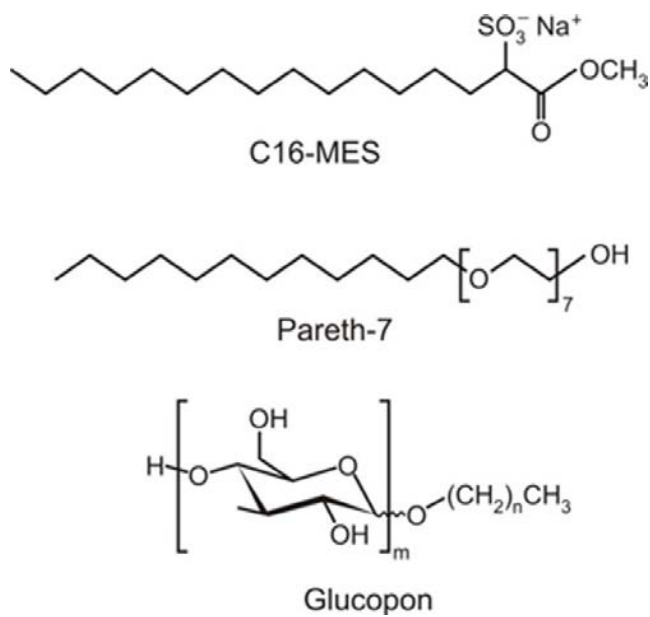

**Fig. 1**. Structural formulae of C16-MES, Pareth-7, and Glucopon.

The aqueous solutions were prepared with deionized water (Elix 3 purification system, Millipore, USA). The concentrated surfactant solutions were mixed and stirred at 40 °C for 1 h for the better solubility of all components. The prepared solution at their natural pH were cooled down and placed in a thermostat at 5 °C at least for 24 h for equilibration. For the long-term storage (at least 3 months), all solutions were kept in a constant climate chamber (Binder KBF-S240) at 5 °C. The absorbance of light was measured by a spectrometer (Jasco V-730) at wavelength $\lambda$ = 500 nm [30,31]. The apparatus detects the ratio between the intensities $I_0$ and $I$ of the incident and transmitted beams, respectively, in terms of $\log_{10}(I_0/I)$. The turbidity is due to light scattering by MES crystallites. Before each absorbance measurement, the flask with the probe was shaken to disperse the available precipitates, if any. The concentrated micellar solutions of the nonionic surfactants (without added MES) are transparent at 5 °C. For the solubility test in the presence of MES, the respective absorbance of the concentrated nonionic surfactant solution without added MES is used for a baseline. The obtained data are summarized in Section 3, where we determined the solubility limits of the studied MES in pure water and their saturation concentrations in micellar solutions of Pareth-7 and Glucopon.

To check experimentally the calculated positions of the phase boundaries, the electrolytic conductivity of solutions at fixed MES mole fractions were carried out by Hanna EC215 conductivity meter.

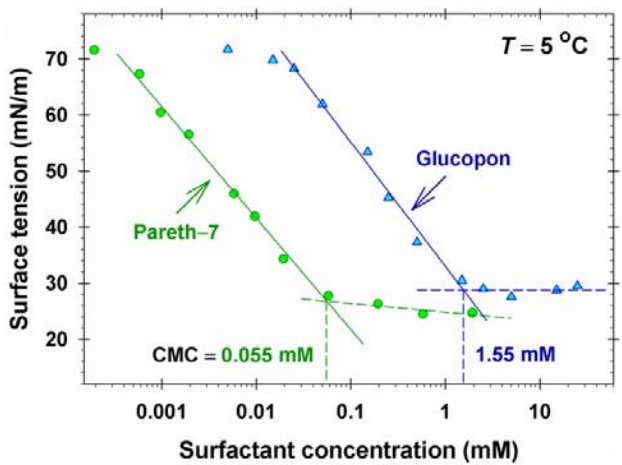

**Fig. 2**. Surface tension isotherms of Pareth-7 and Glucopon measured at 5 °C. The critical micelle concentrations are 0.055 mM and 1.55 mM, respectively.

To obtain the critical micelle concentrations of the nonionic surfactants, we measured their surface tension isotherms at 5 °C by force tensiometer K100 (Krüss, Germany) using the Du Noüy ring (Fig. 2). The CMC of Glucopon is 1.55 mM, while that of Pareth-7 is considerably lower (0.055 mM) because of the longer Pareth-7 hydrophobic tail. The adsorptions at the CMC for Glucopon and Pareth-7 are 4.16 and 3.73 μmol/m², respectively. The larger area per molecule at the CMC of Pareth-7 (44.5 Å²) compared to that of Glucopon (39.9 Å²) is because of the larger size of 7 ethoxy groups compared to 1.2-1.5 glucoside groups. Pareth-7 and Glucopon are nonionic surfactants and the small amount of surface active admixtures does not affect the obtained values of the critical micelle concentrations.

### 3. Experimental results

*3.1. Solubility limits of MES in pure water*

Figs. 3a–3c show the dependence of the light absorbance of MES aqueous solutions on the surfactant concentration. The absorbance of each solution does not change after the first day and remains constant for a long-term storage. In each separate curve, an abrupt increase in the absorbance is observed above a certain concentration. In the case of C14-MES and C16-MES (Fig. 3a), these concentrations are $S_{14}$ = 3.00 mM and $S_{16}$ = 0.695 mM, respectively

(Table 1). The obtained solubility limits, $S_{14}$ and $S_{16}$, are close to the literature data of 3.17 and 0.635 mM [29]. If one plots the absorbance of mixed C16/C18 (80/20) solutions as a function of the concentration of C16-MES, one obtains the same dependence on the concentration as that for the absorbance of C16-MES alone (Fig. 3b). Hence 20 wt% C18-MES in the mixture does not affect the appearance of C16-MES crystallites.

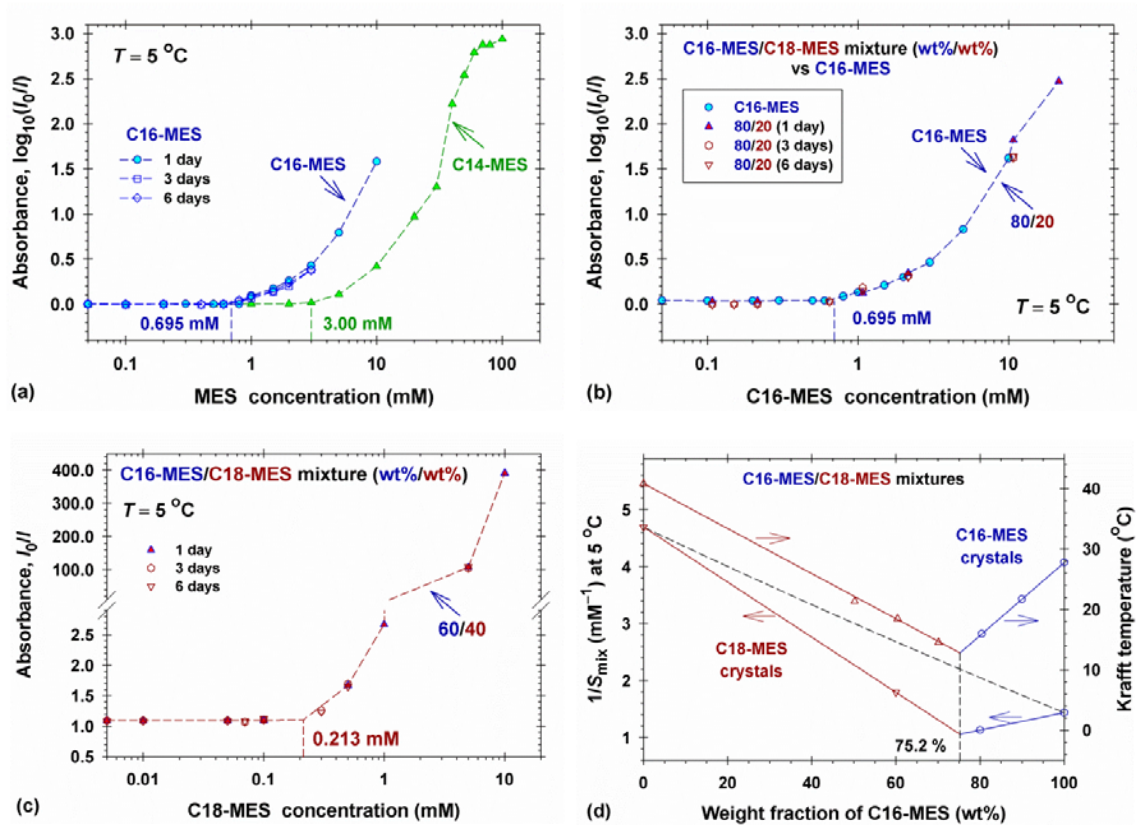

**Fig. 3**. Absorbance and solubility of MES aqueous solutions: a) absorbance of C14-MES and C16-MES; b) comparison between absorbance of C16-MES and that of mixed C16/C18 (80/20) plotted as a function of C16-MES concentration; c) dependence of the absorbance of C16/C18 (60/40) solutions on C18-MES concentration; d) dependence of the Krafft temperature [29] and the solubility limits at 5 °C of C16/C18 mixed aqueous solutions on the weight fraction of C16-MES in MES mixture.

**Table 1**. Solubility limits of C$n$-MES at 5 °C.

| Surfactant: | C12-MES | C14-MES | C16-MES | C18-MES |
|---|---|---|---|---|
| $S_n$ (mM) | 10.7 | 3.00 | 0.695 | 0.213 |

The Krafft temperature of mixed C16-MES/C18-MES aqueous solutions is measured in the literature [29]. If one plots the experimental data (Fig. 3d) as a function of the weight fraction of C16-MES, $w_{16}$, one sees two well pronounced linear trends. With the increase of the weight fraction of C16-MES, $w_{16} \leq 75.2\%$, the Krafft temperature, $T_K$, decreases down to 13 °C. The subsequent increase of $w_{16}$ leads to a linear increase of $T_K$. Hence C16-MES and C18-MES do not form mixed crystals: C18-MES crystals appear for $w_{16} < 75.2\%$; C16-MES crystals are formed for $w_{16} > 75.2\%$. For that reason, the experimental curves in Fig. 3b coincide. From the plot of the absorbance of C16/C18 (60/40) aqueous solutions versus the C18-MES concentration (Fig. 3c), we measured the solubility limit of C18-MES, $S_{18}$, because the concentration of C16-MES is lower than $S_{16}$ (Table 1).

Fig. 3 illustrates the main idea to use C16/C18 (80/20) and (60/40) mixtures. Relatively pure C18-MES is difficult to be synthesized. The C16/C18 (80/20) mixture is chosen to check the hypothesis that C16-MES and C18-MES molecules prefer to form individual crystals, instead of the mixed ones. Respectively, the obtained results for C16/C18 (60/40) mixture give possibility to obtain the solubility limit of C18-MES, $S_{18}$.

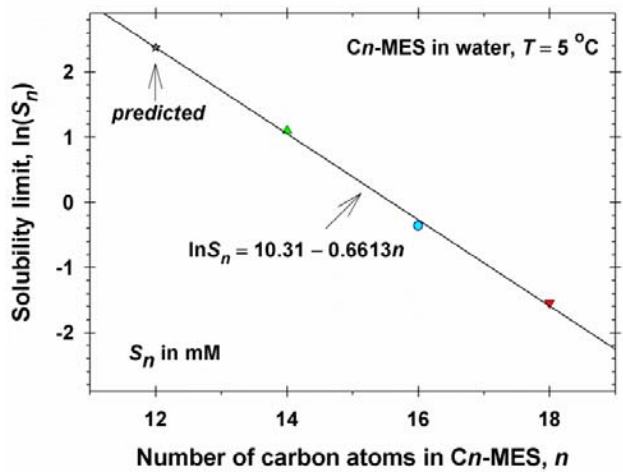

**Fig. 4**. Dependence of the solubility of C$n$-MES in water at 5 °C on the number of carbon atoms, $n$.

From the thermodynamics of ideal mixtures, it follows that the value of the solubility limit for ideal mixed crystal phases, $S_{mix}$, obeys the simple relationship:

$$\frac{1}{S_{mix}} = \frac{x_{16}}{S_{16}} + \frac{x_{18}}{S_{18}} \tag{3.1}$$

where the mole fraction of C16-MES is $x_{16}$ and that of C18-MES is $x_{18}$. The dashed line between points $1/S_{16}$ and $1/S_{18}$ in Fig. 3d is drawn using Eq. (3.1). The experimental points deviate considerably from this line, which is an additional indication that C16-MES and C18-MES molecules prefer to form individual crystals, instead of the mixed ones.

Fig. 4 shows the experimental dependence of the solubility of C$n$-MES in pure water at 5 $^o$C on the number of carbon atoms, $n$. It is well illustrated that ln$S_n$ is a linear function of the number of carbon atoms, $n$, and $\ln S_n = 10.31 - 0.6613n$. The linear regression coefficient in Fig. 4 is 0.9991 and the precision of the slope and intercept are −0.6613±0.02 and 10.31±0.3, respectively. This interpolation dependence allows predicting of the solubility limit of C12-MES in pure water, $S_{12}$ (see Table 1).

*3.2. Saturation concentrations of MES in Glucopon and Pareth-7 micellar solutions*

In the presence of MES and nonionic surfactant with concentration above the CMC, the mixed micelles are formed. At fixed concentration of Glucopon (50, 100, 200, and 300 mM) and Pareth-7 (100, 200, and 300 mM), the capacity of nonionic micelles to incorporate MES is limited. Hence above a given input MES concentration, the unincorporated MES in nonionic micelles will form precipitates and the micellar solutions become turbid. This threshold total MES concentration is called the micellar saturation concentration, $C_{A,sat}$ [30,31]. In the presence of mixed micelles, the turbidity of the solutions changes rather slowly over the storage time (Figs. S1 and S2). In our case three weeks are needed to reach the equilibrium distribution of MES molecules in the micellar, free surfactant in water and crystal phases. In all experiments the storage time was 90 days and after the third week, the turbidity remains constant and does not change at least for three months. The nonionic cosurfactants affect considerably the kinetics of MES crystals formation. The step increase in the absorbance is well pronounced and the saturation concentration is measured with a good precision. Even for the lowest experimental concentration of Glucopon (50 mM), the saturation concentration of MES is about 30 times larger than the respective solubility limit of MES in pure water. With the increase of the nonionic surfactant concentration, the MES saturation concentration increases as well.

To characterize the micellar solutions, we used the following strategy for data processing. In the case of added C16-MES, the mixed solutions contain two components: C16-MES and nonionic surfactants. At the MES saturation concentration, $C_{A,sat}$ ($C_{n,sat}$), the concentration of MES molecules in the form of free monomers, no matter the total MES concentration, is equal to $S_n$. The conditions for the chemical equilibrium between the molecules in the micellar phase and those in the form of free monomers for ideal mixing read [30–32]:

$$\ln S_n = \ln K_{n,\text{mic}} + \ln y_{n,\text{sat}}, \quad \ln c_S = \ln K_{S,\text{mic}} + \ln(1 - y_{n,\text{sat}}) \quad (3.2)$$

Here: $K_{n,\text{mic}}$ and $K_{S,\text{mic}}$ are the respective micellar constant of C$n$-MES and nonionic surfactant molecules in the micellar phase; $y_{n,\text{sat}}$ is the mole fraction of C$n$-MES in the mixed micelles at saturation; $c_S$ is the concentration of free monomers of nonionic surfactant. The conservation of mass equations for the species are [30–32]:

$$S_n + y_{n,\text{sat}} c_{\text{mic}} = C_{n,\text{sat}}, \quad c_S + (1 - y_{n,\text{sat}}) c_{\text{mic}} = C_S \quad (3.3)$$

where $C_S$ is the input nonionic surfactant concentration, $C_T = C_S + C_{n,\text{sat}}$ is the total concentration, and $c_{\text{mic}}$ is the total number of molecules (in mole units) incorporated in the micelles divided by the solution volume. From Eqs. (3.2) and (3.3), one obtains the following relationship between $C_{n,\text{sat}}$ and $C_S$:

$$C_{n,\text{sat}} = S_n - y_{n,\text{sat}} K_{S,\text{mic}} + \frac{y_{n,\text{sat}}}{1 - y_{n,\text{sat}}} C_S \quad (3.4)$$

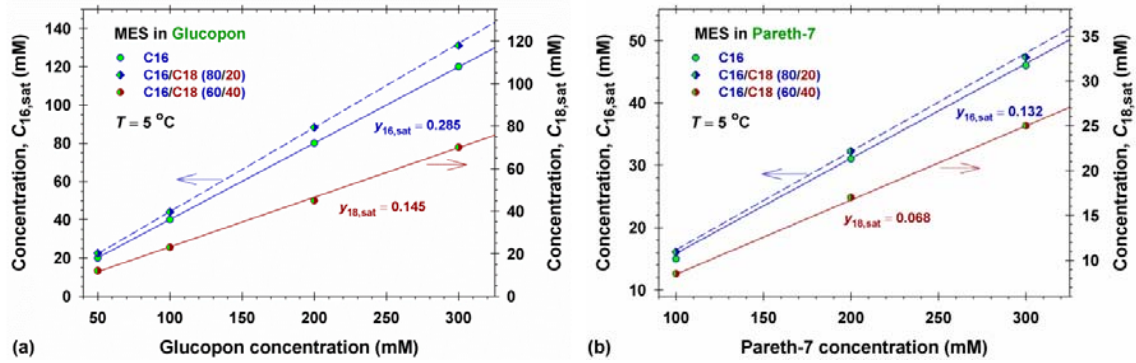

**Fig. 5**. Saturation concentrations of MES in nonionic surfactant micellar solutions. Saturation concentrations, $C_{16,\text{sat}}$ and $C_{18,\text{sat}}$, versus the cosurfactant concentration, $C_S$, for studied MES in (a) Glucopon and (b) Pareth-7 micellar solutions.

Eq. (3.4) shows that the experimental dependence, $C_{n,\text{sat}}$ versus $C_S$, defines the saturation mole fraction, $y_{n,\text{sat}}$, and subsequently $K_{n,\text{mic}} = S_n y_{n,\text{sat}}$, see Eq. (3.2). In Figs. 5a and 5b we plotted the experimental data for $C_{16,\text{sat}}$ versus $C_S$ in the case of C16-MES added to the nonionic micellar solutions. The nonlinear regression analysis in accordance with Eq. (3.4) gives the most probable values of $y_{16,\text{sat}}$, which are equal to 0.285±0.001 and to 0.132±0.002 for Glucopon and Pareth-7 micelles, respectively (solid lines in Figs. 5a and 5b). Note, that $C_{n,\text{sat}}$ is a linear function of $C_S$, see Eq. (3.4), but the intercept, $S_n - y_{n,\text{sat}} K_{S,\text{mic}}$, and the slope, $y_{n,\text{sat}}/(1 - y_{n,\text{sat}})$, are not independent adjustable parameters – they depend only on the saturation mole fraction, $y_{n,\text{sat}}$. The obtained values of $K_{S,\text{mic}}$ from the measured critical micelle concentrations (see Fig. 2) and $K_{16,\text{mic}}$ are summarized in Table 2.

**Table 2**. Micellar constants of C$n$-MES, Glucopon, and Pareth-7 at 5 °C.

|  | $K_{S,mic}$ (mM) | $y_{16,sat}$ | $K_{16,mic}$ (mM) | $y_{18,sat}$ | $K_{18,mic}$ (mM) |
|---|---|---|---|---|---|
| Glucopon | 1.55 | 0.285±0.001 | 2.44 | 0.145±0.003 | 1.47 |
| Pareth-7 | 0.055 | 0.132±0.002 | 5.27 | 0.068±0.003 | 3.13 |

In the case of C16-MES and C18-MES mixtures, one kind of MES crystals appears the first. The observed saturation concentration can be because of the appearance of C16-MES or C18-MES crystals. We denote the component, which first forms precipitates, by A1 and the next one – by A2. Hence the equations of the chemical equilibrium between the molecules in the micellar phase and those in the form of free monomers at the saturation concentration are:

$$\ln S_{A1} = \ln K_{A1,mic} + \ln y_{A1,sat}, \quad \ln c_{A2} = \ln K_{A2,mic} + \ln y_{A2} \tag{3.5a}$$

$$\ln c_S = \ln K_{S,mic} + \ln y_S \tag{3.5b}$$

Here: $K_{A1,mic}$, $K_{A2,mic}$, and $K_{S,mic}$ are the micellar constants of the respective components; $y_{A1,sat}$, $y_{A2}$, and $y_S$ are their mole fractions in the micelles; $y_{A1,sat} + y_{A2} + y_S = 1$; $c_{A2}$ and $c_S$ are the respective free monomer concentrations in the bulk phase. If the total input concentrations are $C_{A1,sat}$, $C_{A2}$, and $C_S$, then the mass balance equations read:

$$S_{A1} + y_{A1,sat} c_{mic} = C_{A1,sat}, \quad c_{A2} + y_{A2} c_{mic} = C_{A2} \tag{3.6a}$$

$$c_S + y_S c_{mic} = C_S \tag{3.6b}$$

One eliminates $c_{A2}$, $y_{A2}$, $c_S$, $y_S$, and $c_{mic}$ from Eqs. (3.5) and (3.6) and obtains the following general relationship at the first saturation point:

$$\frac{y_{A1,sat} C_{A2}}{y_{A1,sat} K_{A2,mic} + C_{A1,sat} - S_{A1}} + \frac{y_{A1,sat} C_S}{y_{A1,sat} K_{S,mic} + C_{A1,sat} - S_{A1}} + y_{A1,sat} = 1 \tag{3.7}$$

Eq. (3.7) for $C_{A2} = 0$ is reduced to the obtained result for two component systems, Eq. (3.4).

The micellar constant, $K_{16,mic}$, is already determined from the experiments with C16-MES (Figs. 5a and 5b, Table 2). The turbidity experiments with C16/C18 (60/40) in pure water (Fig. 3c and 3d) showed that C18-MES precipitates first appear. We assume that in the case of C16/C18 (60/40) MES incorporation in nonionic micelles again the first precipitate, which is responsible for the step increase of the absorbance and turbidity, is that of C18-MES. Hence: A1 is C18-MES; A2 is C16-MES; $C_{A2} = 1.613 C_{A1,sat}$ (see Supplementary material); $K_{A2,mic}$ is $K_{16,mic}$ (Table 2); Eq. (3.7) for a given value of $y_{18,sat}$ is a quadratic equation for the calculation of the saturation concentration, $C_{18,sat}$, which is measured. We fitted experimental data for C16/C18 (60/40) MES + nonionic surfactants systems (Figs. 5a and 5b) using the nonlinear regression model, $C_{18,sat}(y_{18,sat},C_S)$, with only one parameter, $y_{18,sat}$, given by the solution of Eq. (3.7) and obtained $y_{18,sat} = 0.145\pm0.003$ in the case of Glucopon and $y_{18,sat} = 0.068\pm0.003$ for Pareth-7 (solid lines therein). From the saturation mole fraction, $y_{18,sat}$, one calculates the micellar constants, $K_{18,mic} = S_{18}/y_{18,sat}$ (see Tables 1 and 2). As must be true, Fig. S3 in the Supplementary material shows that the calculated from Eqs. (3.5) and (3.6) values of the free C16-MES monomer concentrations, $c_{16}$, are less than the solubility of C16-MES, $S_{16}$, for all studied C16/C18 (60/40) MES + nonionic surfactants systems.

To prove the validity of our assumption, we consider the experimental data for MES saturation concentrations in the case of C16/C18 (80/20). Because of the large enough weight fraction of C16-CMES (see Section 5), the abrupt increase in absorbance versus MES concentration curve corresponds to the saturation concentration of C16-MES. Hence: A1 is C16-MES; A2 is C18-MES; $K_{A2,mic}$ is $K_{18,mic}$ (Table 2); $y_{A1,sat}$ is $y_{16,sat}$ (Table 2); Eq. (3.7) predicts the saturation concentration, $C_{16,sat}$. There are no unknown parameters and one can draw the dashed lines in Figs. 5a and 5b. The perfect agreement between the experimental values of $C_{16,sat}$ and the predicted saturation concentrations manifests the validity of the obtained micellar parameters and the ideality of the mixing of all components in micelles. The calculated values of the monomer concentrations, $c_{18}$, are shown in Fig. S4. For all studied C16/C18 (80/20) MES + nonionic surfactants systems, we obtained that $c_{18} < S_{18}$ and C18-MES does not form crystals in these micellar solutions.

To check the reversibility of crystal formation, all solutions (transparent and turbid) were heated up to 40 $^{\circ}$C for one hour and subsequently cooled down to 25 $^{\circ}$C. As a result all solutions become transparent. We repeated the experimental procedure described in Section 2 using these solutions and obtained the same values of the saturation concentration, $C_{A,sat}$, given in Tables S1 and S2. Hence, the obtained physicochemical parameters in Tables 1 and 2 correspond to the equilibrium reversible thermodynamic processes.

## 4. Phase diagrams for C16-MES and C18-MES in nonionic micellar surfactant solutions

*4.1. Theoretical model and construction of phase diagrams*

The theoretical model for the phase diagrams for fatty acids and alcohols in micellar surfactant solutions is reported in Refs. [30,31]. Below we apply this model to construct the respective phase diagrams for C16-MES and C18-MES. The phase diagrams have four domains (Fig. 6). In the "molecular solution" domain, the solution contains only monomers of components A (C16-MES or C18-MES) and S (nonionic surfactant). In the region "crystals",

MES-crystals coexist with the monomers of components A and S. Oppositely in the domain "micelles", mixed micelles and monomers are present (without MES crystals). Finally, the domain, in which all of the forms (monomers, mixed micelles and MES precipitates) coexist, is denoted as "micelles + MES-crystals". The boundaries between these domains define four phase separation lines, which intersect into quadruple point Q. It is convenient to plot the phase diagrams in terms of the total input species concentration, $C_T$, and the input mole fraction of MES, $z_A = C_A/C_T$. Below we briefly summarize the model equations for the phase separation lines.

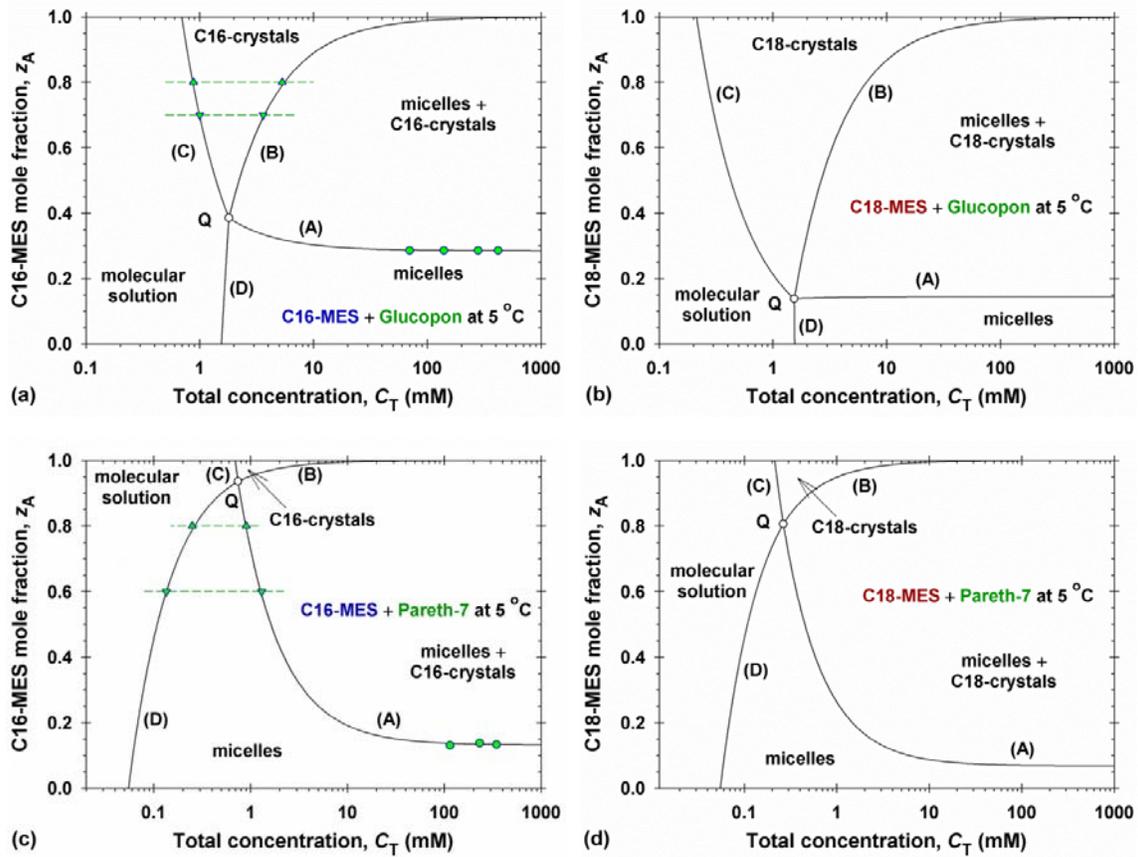

**Fig. 6**. Phase diagrams of C16-MES in (a) Glucopon and (c) Pareth-7 micellar solutions. Phase diagrams of C18-MES in (b) Glucopon and (d) Pareth-7 micellar solutions. The symbols are experimental points, which were measured to verify the phase diagram.

The D-line, (molecular solution)/micelles, describes the critical micelle concentration of mixtures, $CMC_M$, and $C_T = CMC_M$:

$$C_T = CMC_M = y_A K_{A,mic} + (1 - y_A) K_{S,mic}, \quad z_A = y_A K_{A,mic} / C_T \qquad (4.1)$$

where the mole fraction, $y_A$, varies from 0 to $y_{A,sat}$ (Table 2). For $y_A = 0$, one obtains the CMC of the nonionic surfactant. The coordinates $(C_{T,Q}, z_{A,Q})$ of quadruple point Q correspond to the MES saturation mole fraction, that is $y_A = y_{A,sat}$ in Eq. (4.1). The total concentration, $C_{T,Q}$, defines the CMC of the solution at MES saturation, $CMC_{M,sat}$.

Along the C-line, (molecular solution)/crystals, the MES concentration, $c_A = C_A$, is equal to the solubility limit in pure water, $S_A$ (Table 1):

$$z_A = S_A / C_T \quad \text{for} \quad S_A \leq C_T \leq C_{T,Q} \tag{4.2}$$

Along the A-line, micelles/(micelles+MES-crystals), the free MES monomer concentration is equal to the solubility limit, $S_A$ (Table 1), and the micellar mole fraction of MES is equal to the saturation one, $y_{A,sat}$ (Table 2). Hence, Eq. (3.4) is represented in terms of $C_T$ as follows:

$$z_A = y_{A,sat} + \frac{y_{A,sat}}{C_T}(1 - y_{A,sat})(K_{A,mic} - K_{S,mic}) \tag{4.3}$$

for $C_T \geq CMC_{M,sat}$. In fact, our experimental data for C16-MES lie on this phase separation line, see the symbols (circles) in Figs. 6a and 6c.

The B-line separates the MES crystals and the micelles + MES crystals phases. Along this line, the micellar mole fractions of components are $y_A = y_{A,sat}$ and $y_S = 1 - y_{A,sat}$. Thus the respective nonionic surfactant concentration is $C_S = (1 - y_{A,sat})K_{S,mic}$ and the MES concentration is $S_A$. As a result:

$$C_T = (1 - y_{A,sat})K_{S,mic} + S_A \quad \Rightarrow \quad z_A = 1 - \frac{CMC_{M,sat} - S_A}{C_T} \tag{4.4}$$

for $C_T \geq CMC_{M,sat}$.

The calculated phase diagrams for individual C16-MES and C18-MES in Glucopon and Pareth-7 solutions are shown in Fig. 6. The following general conclusions can be drawn. Because of the lower solubility of C18-MES in water, the regions with turbid solutions are wider than those for C16-MES. Note that this conclusion is not trivial, because $K_{16,mic} > K_{18,mic}$ so that for equal values of $y_A$, the monomer concentration of C16-MES is larger than that of C18-MES but the solubility of C16-MES is also larger. The ratios, $K_{16,mic}/K_{18,mic}$, are equal to 1.66 and 1.68 for Glucopon and Pareth-7 (Table 2), respectively, while the ratio between $S_{16}/S_{18}$ is 3.26 (Table 1). Hence the effect of the larger solubility of C16-MES dominates. The saturation micellar mole fractions of MES in Pareth-7 are considerably lower than those in Glucopon (Table 2), i.e. Pareth-7 is unable to incorporate as many MES molecules in micelles. This corresponds to wider turbidity regions and narrower regions with MES crystals without micelles in the case of Pareth-7 compared to Glucopon.

*4.2. Verification of the phase diagram*

The phase diagrams can be verified by measuring properties of the surfactant solutions, which exhibit kinks or jumps when crossing the boundaries between the phase domains. For example, the experimental points (circles) on the A-line in Figs. 6a and 6c correspond to the kinks of absorbance.

To check also the B and C lines in Fig. 6a, we measured the electrolytic conductivity of the respective surfactant solutions at fixed $z_A$ = 0.8 and 0.7 (two horizontal cross-sections of the phase diagram). The B-line of a phase diagram (Fig. 6a) separates the domains with crystallites and crystallites + micelles. The appearance/disappearance of the mixed micelles (which contain anionic surfactant) at the B-line leads to a kink in conductivity (Supplementary material, Fig. S5). Likewise, the disappearance of the crystallites of the ionic surfactant (which maintain almost constant the bulk concentration of this component) at the C-line leads to another kink in conductivity (Fig. S5). The observed kinks in conductivity (triangles in Fig. 6a) are in a very good agreement with the calculated B and C lines of the phase diagram.

The points (triangles) in Fig. 6c are experimental data obtained in a similar way (for details, see the Supplementary material, Figs. S6 and S7). In this case, the experimental points are again in an excellent agreement with the calculated phase boundary lines.

## 5. Phase diagrams for three component micellar surfactant solutions

The general phase diagrams for two partially soluble components, A1 and A2, in micellar surfactant, S, solutions have complex shapes. The simplest case is the following: components A1 and A2 cannot form mixed precipitates; the mixing of the three components in micelles is ideal with micellization constants $K_{A1,mic}$, $K_{A2,mic}$, and $K_{S,mic}$, respectively. The solubility limits of components A1 and A2 in pure water are denoted by $S_{A1}$ and $S_{A2}$. The input concentrations of species are $C_{A1}$, $C_{A2}$, and $C_S$.

In terms of the total concentration, $C_T = C_{A1} + C_{A2} + C_S$, and the mole fractions of the partially soluble components, $z_{A1} = C_{A1}/C_T$ and $z_{A2} = C_{A2}/C_T$, the phase diagrams are 3D and their visualization is qualitative. For that reason, it is convenient to use 2D quantitative plot of the cross-sections at given molar (weight) ratio between components A1 and A2. Hence the respective cross-section can be presented in terms of the total input mole fraction of partially soluble components, $z_A = z_{A1} + z_{A2}$, and total concentration $C_T$.

We define the molar ratio of the two components in MES mixture, $\lambda$:

$$\lambda \equiv \frac{C_{A1}}{C_{A2}} = \frac{z_{A1}}{z_{A2}} \ , \quad z_{A1} = \frac{\lambda z_A}{1+\lambda} \ , \quad z_{A2} = \frac{z_A}{1+\lambda} \tag{5.1}$$

At the D-line, (molecular solution)/micelles, all components are in a molecular form (Figs. 7 and 6). The total concentration is equal to the CMC of the nonionic surfactant for $z_A = 0$. With the increase of $z_A$, the concentration of one of the components in the MES mixture reaches the solubility limit. Thus the simple rule for the ordering of MES components is the following. The first MES component is that, which satisfies the inequality:

$$\lambda = \frac{C_{A1}}{C_{A2}} > \frac{S_{A1}}{S_{A2}} \tag{5.2}$$

From Eq. (5.2), one obtains that: a) for the mixture C16/C18 (80/20) MES + nonionic surfactant, the first component is C16-MES; b) for the mixture C16/C18 (60/40) MES + nonionic surfactant, the first component is C18-MES, see Supplementary material.

*5.1. Phase separation lines and quadruple points*

*Line D: (molecular solution)/micelles boundary.* At this phase separation boundary all components are in a molecular form and the conditions for the chemical equilibrium between micellar and bulk phases read:

$$C_S = K_{S,mic} y_S \ , \quad C_{A1} = K_{A1,mic} y_{A1} \leq S_{A1} \ , \quad C_{A2} = K_{A2,mic} y_{A2} < S_{A2} \tag{5.3}$$

where $0 \leq y_{A1} \leq y_{A1,sat}$. From Eqs. (5.1) and (5.3), one obtains the following expressions for $y_S$ and $y_{A2}$:

$$y_{A2} = \frac{K_{A1,mic} y_{A1}}{K_{A2,mic} \lambda} \ , \quad y_S = 1 - y_{A1} - \frac{K_{A1,mic} y_{A1}}{K_{A2,mic} \lambda} \tag{5.4}$$

Eqs. (5.1), (5.3) and (5.4) suggest describing line D in a parametric form with respect to $y_{A1}$. The expression for the critical micelle concentration of the mixture, $C_T = CMC_M$, is:

$$C_T(y_{A1}) = K_{S,mic} + (K_{A1,mic} - K_{S,mic}) y_{A1} + (K_{A2,mic} - K_{S,mic}) \frac{K_{A1,mic} y_{A1}}{K_{A2,mic} \lambda} \tag{5.5a}$$

and that for the input MES mole fraction, $z_A$, reads:

$$z_A(y_{A1}) = \frac{K_{A1,mic}(1+\lambda)}{C_T \lambda} y_{A1} \tag{5.5b}$$

where $0 \leq y_{A1} \leq y_{A1,sat}$. One sees that in the case of one MES component, A1, one has $\lambda \to \infty$ and Eqs. (5.5a) and (5.5b) are reduced to Eq. (4.1).

This phase separation line (Figs. 7 and 8) finishes at the first quadruple point, Q1, at which the A1 precipitate appears and $y_{A1} = y_{A1,sat}$. Hence coordinates $(C_{T,Q1}, z_{A,Q1})$ of point Q1 are calculated from Eqs. (5.5a) and (5.5b), in which one substitutes $y_{A1} = y_{A1,sat}$.

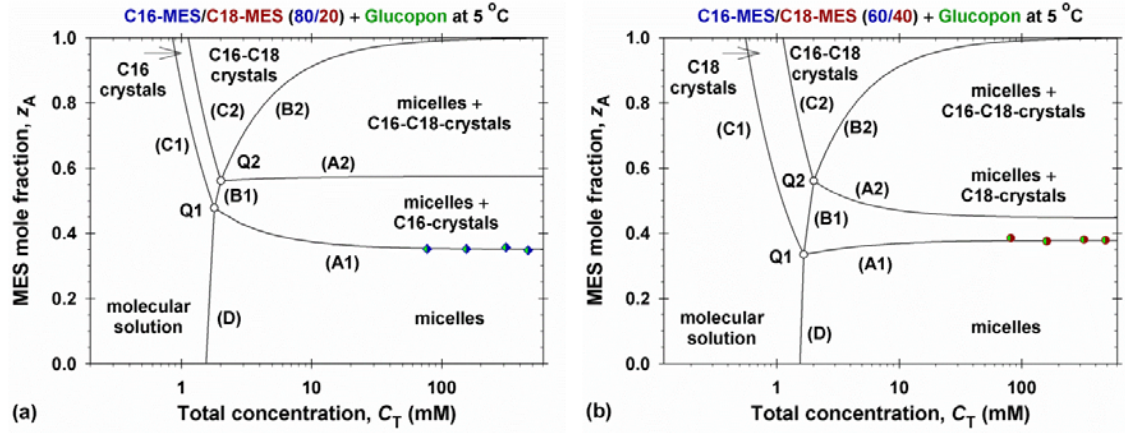

**Fig. 7**. Phase diagrams of (a) C16/C18 (80/20) MES and (b) C16/C18 (60/40) MES in Glucopon micellar solutions.

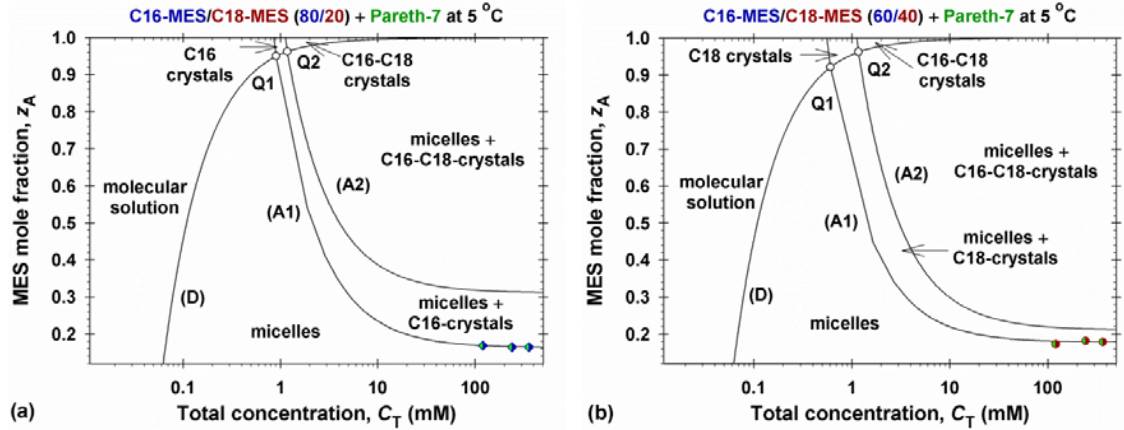

**Fig. 8**. Phase diagrams of (a) C16/C18 (80/20) MES and (b) C16/C18 (60/40) MES in Pareth-7 micellar solutions.

*Line C1: (molecular solution)/(A1-crystals) boundary*. The A1-crystals appear the first and the concentration of A1 is equal to the solubility limit, $C_{A1} = S_{A1}$ and $z_{A1} = S_{A1}/C_T$, at line C1 (Figs. 7 and 8). Using the definition, Eq. (5.1), one arrives to the equation describing this phase separation line:

$$z_A = \frac{1+\lambda}{\lambda}\frac{S_{A1}}{C_T} \quad \text{for} \quad \frac{1+\lambda}{\lambda}S_{A1} \leq C_T \leq C_{T,Q1} \tag{5.6}$$

*Line A1: micelles/(micelles + A1-crystals) boundary.* This line separates the domain with micelles and that with micelles and A1-crystals. The A1 phase separation line starts from the first quadruple point, Q1. Note that the number of all molecules incorporated in the micellar phase, $c_{\text{mic}}$, is equal to zero at point Q1. It is convenient to describe line A1 in a parametric form with respect to $c_{\text{mic}}$. The equations describing the chemical equilibrium and mass conservation along line A1 are Eqs. (3.5) and (3.6).

From Eq. (5.1), we represent the mass balance equations, Eq. (3.6), in the following form:

$$C_{A1} = z_{A1}C_T = S_{A1} + y_{A1,\text{sat}}c_{\text{mic}} \quad \Rightarrow \quad z_A C_T = \frac{1+\lambda}{\lambda}(S_{A1} + y_{A1,\text{sat}}c_{\text{mic}}) \tag{5.7a}$$

$$C_{A2} = z_{A2}C_T = c_{A2} + y_{A2}c_{\text{mic}} \quad \Rightarrow \quad z_A C_T = (1+\lambda)(K_{A2,\text{mic}} + c_{\text{mic}})y_{A2} \tag{5.7b}$$

$$C_S = (1-z_A)C_T = c_S + y_S c_{\text{mic}} \quad \Rightarrow \quad (1-z_A)C_T = (K_{S,\text{mic}} + c_{\text{mic}})y_S \tag{5.7c}$$

Using the identity, $y_{A1,\text{sat}} + y_{A2} + y_S = 1$, one eliminates $y_S$ from Eq. (5.7c) and solves the obtained system of equations to derive the parametric representation of the equations describing line A1:

$$y_{A2}(c_{\text{mic}}) = \frac{S_{A1} + y_{A1,\text{sat}}c_{\text{mic}}}{\lambda(K_{A2,\text{mic}} + c_{\text{mic}})} \tag{5.8a}$$

$$C_T(c_{\text{mic}}) = (K_{S,\text{mic}} + c_{\text{mic}})(1 - y_{A1,\text{sat}}) + [(1+\lambda)K_{A2,\text{mic}} - K_{S,\text{mic}} + \lambda c_{\text{mic}}]y_{A2} \tag{5.8b}$$

$$z_A(c_{\text{mic}}) = \frac{1+\lambda}{\lambda}\frac{S_{A1} + y_{A1,\text{sat}}c_{\text{mic}}}{C_T} \tag{5.8c}$$

for $c_{\text{mic}} \geq 0$ (Figs. 7 and 8). For one MES component, A1, $\lambda \to \infty$ and Eqs. (5.8) are simplified to the result given by Eq. (4.3).

*Line B1: A1-crystals/(micelles + A1-crystals) boundary.* For this line: $y_{A1} = y_{A1,\text{sat}}$; the line starts from the first quadruple point, Q1, where $y_{A2} = y_{A2,Q1}$, and ends at the second quadruple point, Q2, where $y_{A2} = y_{A2,\text{sat}}$. Hence the most convenient parametric form of line B1 is $C_T(y_{A2})$ and $z_A(y_{A2})$. From Eq. (5.4) written at the first quadruple point, $y_{A1} = y_{A1,\text{sat}}$, one calculates:

$$y_{A2,Q1} = \frac{K_{A1,\text{mic}}y_{A1,\text{sat}}}{K_{A2,\text{mic}}\lambda} < y_{A2,\text{sat}} \tag{5.9}$$

Note that the inequality, Eq. (5.9), is equivalent to the rule of ordering of components A1 and A2 given by the inequality, Eq. (5.2). At the B1-line, the equations following from the chemical equilibrium between micellar and bulk phases read:

$$C_S = K_{S,\text{mic}}y_S, \quad S_{A1} = K_{A1,\text{mic}}y_{A1,\text{sat}}, \quad C_{A2} = K_{A2,\text{mic}}y_{A2} \leq S_{A2} \tag{5.10}$$

From Eqs. (5.1) and (5.10) one derives:

$$C_S = (1-z_A)C_T = K_{S,mic}(1-y_{A1,sat} - y_{A2}), \quad C_{A2} = \frac{z_A}{1+\lambda}C_T = K_{A2,mic}y_{A2} \tag{5.11}$$

The solution of the system of equations, Eq. (5.11), leads to the parametric definition of the line B1:

$$C_T(y_{A2}) = (1+\lambda)K_{A2,mic}y_{A2} + K_{S,mic}(1-y_{A1,sat} - y_{A2}) \tag{5.12a}$$

$$z_A(y_{A2}) = (1+\lambda)y_{A2}\frac{K_{A2,mic}}{C_T} \tag{5.12b}$$

for $y_{A2,Q1} \leq y_{A2} \leq y_{A2,sat}$. The coordinates $(C_{T,Q2}, z_{A,Q2})$ of the second quadruple point, Q2, are calculated from Eqs. (5.12) with $y_{A2} = y_{A2,sat}$ (Figs. 7 and 8).

The main difference between the phase diagrams of one- and those of two-component partially soluble species, is that for two components, there are domains with two precipitates (A1-crystals and A2-crystals). These crystals can coexist with molecular forms of all species (A1-crystals + A2-crystals domain) or with micelles and free monomers (A1-crystals + A2-crystals + micelles domain). For simplicity, we denote these regions as A1-A2-crystals and micelles+A1-A2-crystals (Figs. 7–9). These two new domains are separated by three additional boundaries in the phase diagrams.

*Line C2: A1-crystals/A1-A2-crystals.* The precipitates of A2 appear at line C2, so that $C_{A2} = S_{A2}$. From Eq. (5.1), one represents concentration $C_{A2}$ in terms of $z_A$ and $C_T$ and obtains: $z_A = (1+\lambda)S_{A2}/C_T$. One sees from Eq. (5.6) that C2-line: is parallel to the C1-line; is shifted to the right with respect to the total concentration. The total concentration, $C_T$, changes from $(1+\lambda)S_{A2}$ to abscise $C_{T,Q2}$ of the quadruple point Q2.

*Line A2: (micelles + A1-crystals)/(micelles + A1-A2-crystals) boundary.* At line A2, the expressions for the chemical equilibrium applied to both components A1 and A2 define their saturation mole fractions in micelles: $S_{A1} = K_{A1,mic}y_{A1,sat}$; $S_{A2} = K_{A2,mic}y_{A2,sat}$. Hence $y_S = 1 - y_{A1,sat} - y_{A2,sat}$. The mass conservation equation for the A2-component gives possibility to obtain the total number of molecules in micelles per unit volume, $c_{mic}$:

$$C_{A2} = S_{A2} + y_{A2,sat}c_{mic} \Rightarrow c_{mic} = \frac{C_{A2} - S_{A2}}{y_{A2,sat}} \tag{5.13a}$$

Respectively, for the nonionic surfactant, one obtains the relationships:

$$c_S = K_{S,mic}y_S, \quad C_S = c_S + y_S c_{mic} = (K_{S,mic} + c_{mic})y_S \tag{5.13b}$$

The elimination of $c_{mic}$ from Eqs. (5.13a) and (5.13b) leads to the following formula:

$$y_{A2,sat}C_S = (y_{A2,sat}K_{S,mic} + C_{A2} - S_{A2})y_S \tag{5.13c}$$

Finally, from Eqs. (5.1) and (5.13c), we obtain the following explicit dependence of $z_A$ on $C_T$:

$$z_A = \frac{(1+\lambda)y_{A2,sat}}{1 - y_{A1,sat} + \lambda y_{A2,sat}} - \frac{(1+\lambda)y_{A2,sat}(1 - y_{A1,sat} - y_{A2,sat})}{1 - y_{A1,sat} + \lambda y_{A2,sat}} \frac{K_{S,mic} - K_{A2,mic}}{C_T} \quad (5.14)$$

for $C_T \geq C_{T,Q2}$.

*Line B2: A1-A2-crystals/(micelles + A1-A2-crystals) boundary.* Along this line, Eq. (5.13b) with $c_{mic} = 0$ is valid and $y_S = 1 - y_{A1,sat} - y_{A2,sat}$. Hence from the definition of parameters, Eq. (5.1), one derives the relationship:

$$z_A = 1 - (1 - y_{A1,sat} - y_{A2,sat})\frac{K_{S,mic}}{C_T} \quad (5.15)$$

valid for $C_T \geq C_{T,Q2}$.

*5.2. Numerical results and discussion*

Following the proposed theoretical model, we calculated the positions of the phase separation lines and quadruple points in the phase diagrams of C16/C18 (80/20) MES and C16/C18 (60/40) MES in Glucopon micellar solutions (Fig. 7). All needed physicochemical parameters are summarized in Tables 1 and 2. The symbols in Fig. 7 correspond to the experimental points from Fig. 5b. One sees that the first precipitate, which is formed for C16/C18 (80/20), is that of C16-crystals (Fig. 7a). With the increase of the amount of added MES, the solutions become more turbid because of the simultaneous appearance of the second (C18-crystals) precipitates. The decrease of C16-MES in the mixture, C16/C18 (60/40), changes the order of the appearance of MES-crystals (Fig. 7b). Because of the lower solubility of C18-MES in pure water, the region of turbid solutions is considerably wider for 60/40 than that for 80/20. For $C_T > 2$ mM, the solutions containing C16/C18 (80/20) are clear with the decrease of $z_A$ from 48% to 35% and the rise of the total concentration, $C_T$. While for C16/C18 (60/40) this increases from 33% to 38% with the rise of $C_T$.

The calculated phase diagrams of MES mixtures in Pareth-7 micellar solutions are shown in Fig. 8. The symbols therein correspond to experimental data from Fig. 5b. From one side, the CMC of Pareth-7 is considerably lower than that of Glucopon and at the same concentrations the Pareth-7 solutions contain more micelles. Nevertheless, the turbidity region in Fig. 8 is wider than that in Fig. 7, because of the lower saturation mole fractions of MES in Pareth-7 micelles (Table 2). The ordering of the domains in phase diagrams is the same as that for Glucopon micellar solutions. The positions of the quadruple points are close to each other, which results in narrow domains with one type of MES crystals. It is interesting to note that for $C_T > 100$ mM, the clear micellar zone is observed for close values of $z_A$, < 16% for C16/C18 (80/20) and < 18% for C16/C18 (60/40). The domains containing C16-C18-crystals are wider for C16/C18 (60/40) compared to those for C16/C18 (80/20).

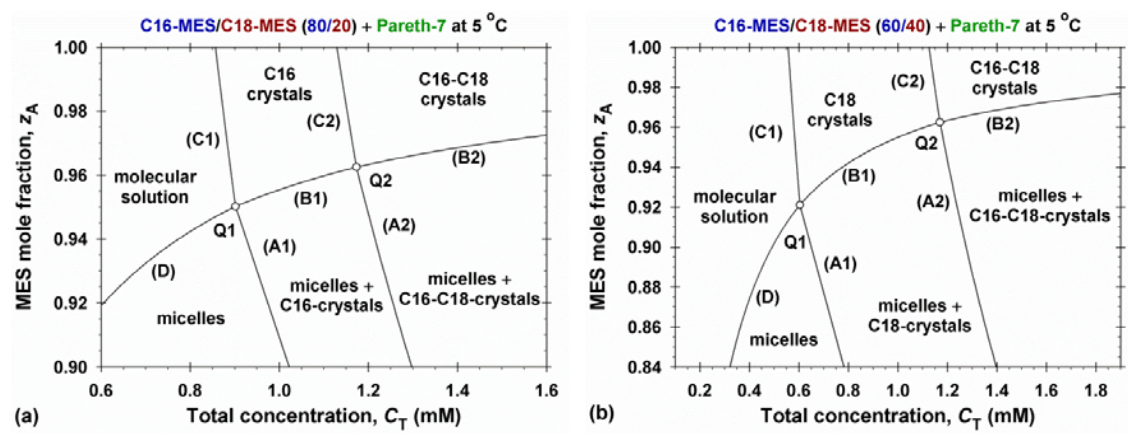

**Fig. 9**. Enlarged view of the domains around the quadruple points of the phase diagrams of (a) C16/C18 (80/20) MES and (b) C16/C18 (60/40) MES in Pareth-7 micellar solutions.

Fig. 9 shows the enlarged view of Fig. 8 around the quadruple points of the respective phase diagrams. The positions of points Q2 in Fig. 9a and 9b are very close to each other with coordinates (1.17, 0.962). Hence the domains containing C16-C18-crystals around quadruple point Q2 practically coincides. The quadruple point, Q1, has coordinates (0.901, 0.950) for C16/C18 (80/20) and (0.604, 0.921) for C16/C18 (60/40). The smaller concentration difference between Q1 and Q2 is another indication that the domains containing individual C16-crystals around point Q1 in Fig. 9a are narrower than those containing individual C18-crystals in Fig. 9b.

## 6. Conclusions

Below the Krafft point the C$n$-MES are partially soluble in pure water and their solubility limits decrease considerably with the decrease of temperature [29]. From the precise measurements of the light adsorption of C$n$-MES and their mixtures in water at 5 °C, we obtained new data for the solubility limits, $S_n$, for $n$ = 14, 16, and 18, Table 1. The logarithm of $S_n$ becomes a linear function of the number of carbon atoms in the alkyl chain, which allows prediction of the solubility limit of C12-MES, Fig. 4. The important experimental observation is that in all mixtures, C16-MES and C18-MES do not form mixed crystals – the first precipitate appears for that MES, for which the respective solubility limit is reached with the increase of concentration. Based on this conclusion, we formulated a simple rule for the ordering of MES precipitates in mixtures, see Eq. (5.2).

An efficient way to increase the solubility of poorly water-soluble drugs, fatty alcohols and acids [30,31,33,34] is to incorporate their molecules in the micelles. Here, the considerable increase of the solubility of MES is realized in the presence of nonionic surfactant micelles because of the effective incorporation of MES molecules in the micellar phase. The capacity of Glucopon and Pareth-7 micelles is characterized by the saturation mole fraction of MES, $y_{n,\text{sat}}$, in micelles. From the experimental data, Fig. 5 and Supplementary material, we obtained the respective micellar constants for C16- and C18-MES in both nonionic surfactant micelles, Table 2. The micelles of Glucopon show about two times greater capacity than those of Pareth-7 (1.97 times for C16-MES and 1.94 – for C18-MES).

The determined complete set of physicochemical parameters enabled us to calculate and construct quantitative phase diagrams for mixed MES and micellar surfactant solution. For one partially soluble component (C16- and C18-MES), these diagrams are calculated in Fig. 6 using the theory developed in Refs. [30,31]. For mixed C16/C18 MES solutions the phase diagrams become 3D and their quantitative representation is possible for a given molar ratio of MES components in the mixture. A new theory for two partially soluble components and one soluble surfactant solutions is proposed in Section 5. The obtained results manifest the formation of six domains in the phase diagram (Figs. 7–9): molecular solutions; molecular solutions with one and two precipitates; micellar solutions coexisting with monomers of components; micellar solutions with one and two precipitates. These domains are separated by seven phase separation lines, which intersect in two quadruple points. The coordinates of all phase separation lines, the CMC of mixtures, and quadruple points are calculated numerically.

The results may contribute to understanding, quantitative interpretation and prediction of phase diagrams of mixed micellar solutions when the Krafft temperatures of respective components are considerably different. The new theoretical approach upgrades the available models in the literature [30,31,35].

**CRediT authorship contribution statement**

**K.D. Danov:** Software, Formal analysis, Supervision, Writing – original draft. **P.A. Kralchevsky:** Conceptualization, Methodology, Supervision. **R.D. Stanimirova, T.G. Slavova, V.I. Yavrukova:** Investigation, Data curation. **Yee Wei Ung:** Project administration, Funding acquisition, Conceptualization, Supervision. **Emily Tan, Hui Xu:** Resources, Data curation. **J.T. Petkov:** Conceptualization, Methodology.

**Declaration of Competing Interest**

The authors declare that they have no known competing financial interests or personal relationships that could have appeared to influence the work reported in this paper.

**Acknowledgements**

The authors gratefully acknowledge the support from KLK OLEO. K. Danov acknowledges the support from the Operational Programme "Science and Education for Smart Growth", Bulgaria, project No. BG05M2OP001-1.002-0023.**Appendix A. Supplementary material**

Supplementary material to this article can be found online at

## Supplementary Material

for the article

**Solubility of ionic surfactants below their Krafft point in mixed micellar solutions: Phase diagrams for methyl ester sulfonates and nonionic cosurfactants**

<u>Authors</u>: Krassimir D. Danov, Rumyana D. Stanimirova, Peter A. Kralchevsky, Tatiana G. Slavova, Veronika I. Yavrukova, Yee Wei Ung, Emily Tan, Hui Xu, Jordan T. Petkov

*C16-MES in 100 mM Glucopon micellar solutions*

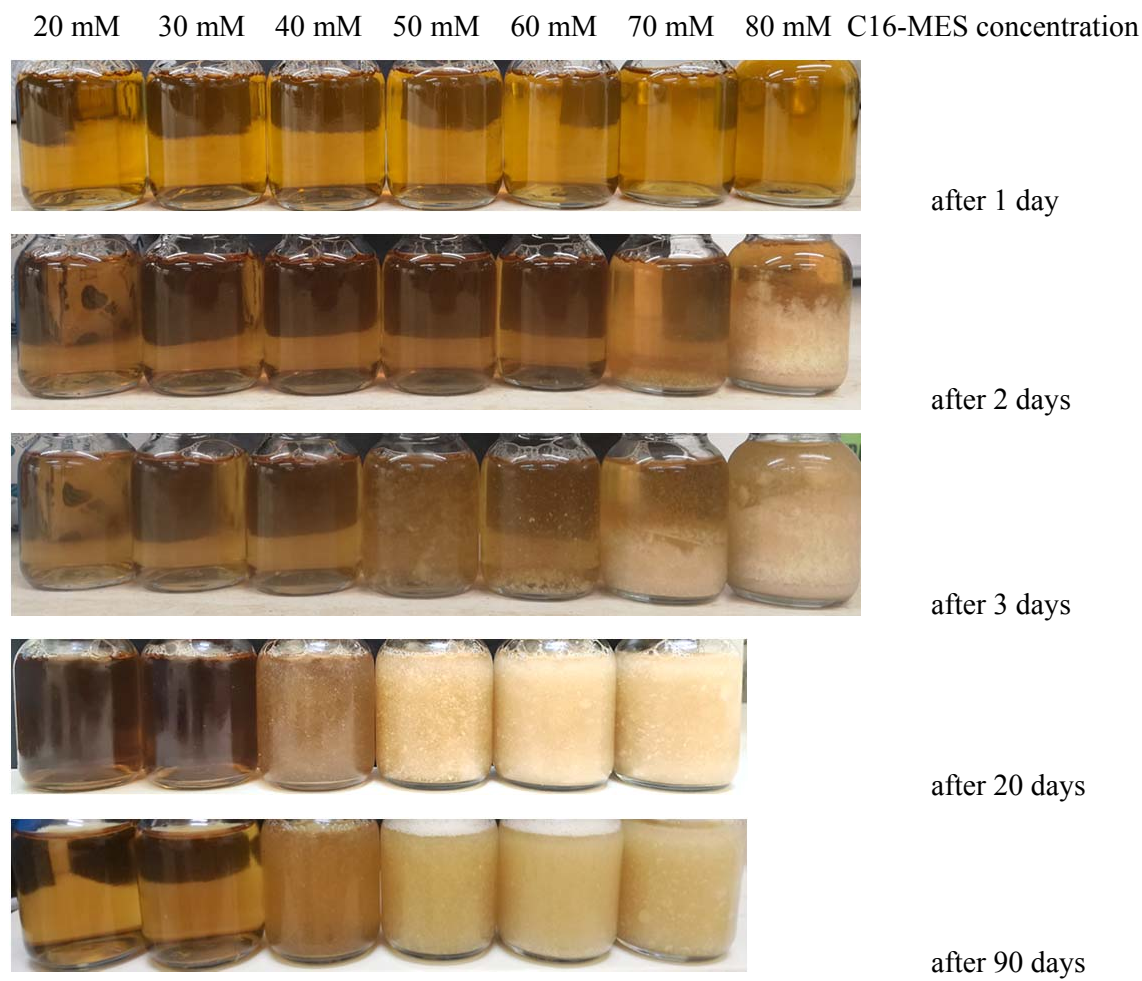

**Fig. S1**. Photographs of C16-MES in 100 mM Glucopon micellar solutions taken after different storage time in the climate chamber at 5 $^\circ$C. The concentrations of C16-MES are shown on the top of the figure. The measurements of the light absorbance show that $C_{16,sat}$ = 40 mM.

*C16-MES in 100 mM Pareth-7 micellar solutions*

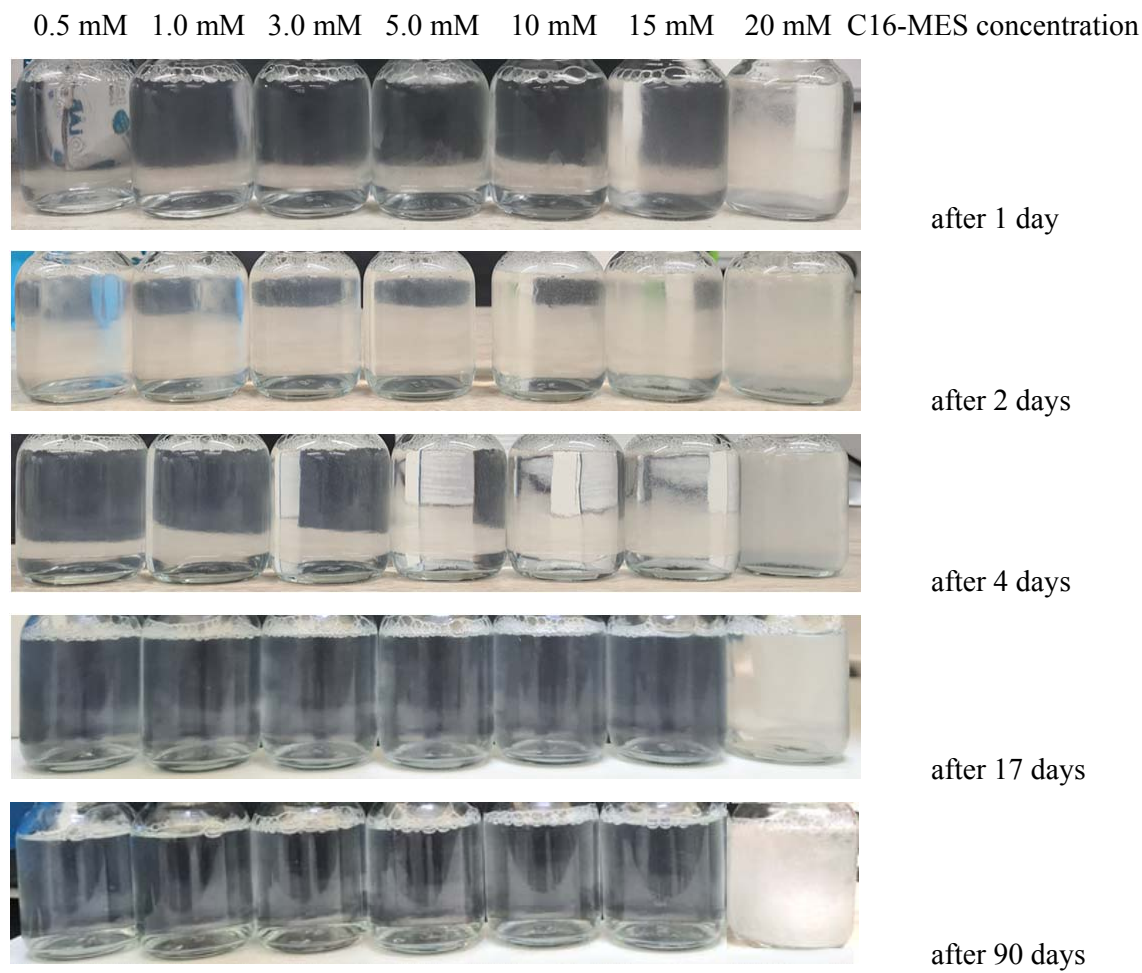

**Fig. S2**. Photographs of C16-MES in 100 mM Pareth-7 micellar solutions taken after different storage time in the climate chamber at 5 °C. The concentrations of C16-MES are shown on the top of the figure. The measurements of the light absorbance show that $C_{16,sat}$ = 15 mM.

Figs. S1 and S2 show photographs of C16-MES in 100 mM Glucopon and C16-MES in 100 mM Pareth-7 micellar solutions taken after different storage time in the climate chamber at 5 °C. It is well illustrated that the turbidity of solutions changes during the first two-three weeks. After the third week, the turbidity remains constant and does not change at least for three months. For all studied concentrations of the nonionic cosurfactants, the equilibration period of solutions was not more than three weeks. The experimental data for the MES saturation concentration, $C_{A,sat}$, are summarized in Tables S1 and S2. These data are illustrated in Figs. 5a and 5b and processed using the algorithms discussed in Section 3.2 to obtain the values of micellar parameters $y_{16,sat}$, $y_{18,sat}$, $K_{16,mic}$, and $K_{18,mic}$ (Table 2).

**Table S1**. Experimental data for the MES saturation concentration, $C_{A,sat}$, in mixed micellar solutions of MES and Glucopon.

| $C_S$ (mM) | C16-MES | C16/C18 (80/20) | | | C16/C18 (60/40) | | |
|---|---|---|---|---|---|---|---|
| | $C_{16,sat}$ (mM) | $C_{A,sat}$ (wt%) | $C_{16,sat}$ (mM) | $C_{18}$ (mM) | $C_{A,sat}$ (wt%) | $C_{16}$ (mM) | $C_{18,sat}$ (mM) |
| 50 | 20 | 1.05 | 22.6 | 5.25 | 1.2 | 19.4 | 12 |
| 100 | 40 | 2.05 | 44.1 | 10.25 | 2.3 | 37.1 | 23 |
| 200 | 80 | 4.1 | 88.2 | 20.5 | 4.5 | 72.6 | 45 |
| 300 | 120 | 6.1 | 131 | 30.5 | 7.0 | 113 | 70 |

**Table S2**. Experimental data for the MES saturation concentration, $C_{A,sat}$, in mixed micellar solutions of MES and Pareth-7.

| $C_S$ (mM) | C16-MES | C16/C18 (80/20) | | | C16/C18 (60/40) | | |
|---|---|---|---|---|---|---|---|
| | $C_{16,sat}$ (mM) | $C_{A,sat}$ (wt%) | $C_{16,sat}$ (mM) | $C_{18}$ (mM) | $C_{A,sat}$ (wt%) | $C_{16}$ (mM) | $C_{18,sat}$ (mM) |
| 100 | 15 | 0.75 | 16.1 | 3.75 | 0.85 | 13.7 | 8.5 |
| 200 | 31 | 1.5 | 32.3 | 7.50 | 1.7 | 27.4 | 17 |
| 300 | 46 | 2.2 | 47.3 | 11.0 | 2.5 | 40.3 | 25 |

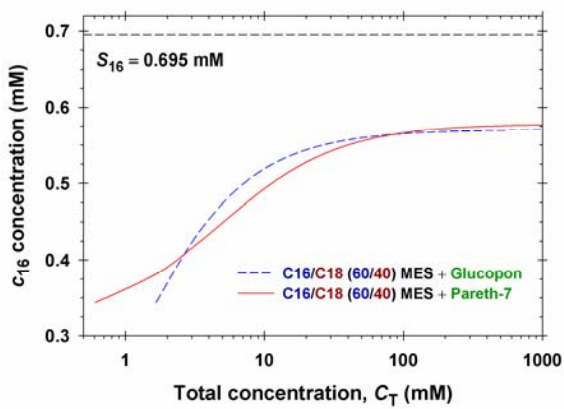

**Fig. S3**. Calculated bulk monomer concentrations of C16-MES, $c_{16}$, corresponding to the saturation concentration, $C_{A,sat}$, for C16/C18 (60/40) MES + nonionic surfactant micellar solutions. $S_{16}$ = 0.695 mM is the solubility limit of C16-MES in water at 5 °C.

In order to obtain the micellar parameters of C18-MES from experimental data (Figs. 5a and 5b), we assumed that the turbidity of C16/C18 (60/40) MES + nonionic surfactant micellar solutions appear because of the C18-MES crystallites. Hence, the respective monomer bulk concentration of C16-MES, $c_{16}$, should be lower than the solubility of C16-MES in pure water, $S_{16}$, measured at 5 °C. Indeed, the respective results from the calculations of $c_{16}$ corresponding to the saturation concentration, $C_{A,sat}$, show that $c_{16} < S_{16}$ (Fig. S3).

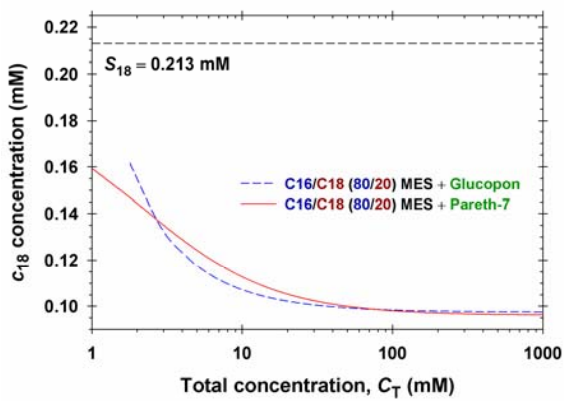

**Fig. S4**. Calculated bulk monomer concentrations of C18-MES, $c_{18}$, corresponding to the saturation concentration, $C_{A,sat}$, for C16/C18 (80/20) MES + nonionic surfactant micellar solutions. $S_{18} = 0.213$ mM is the solubility limit of C18-MES in water at 5 °C.

In contrast for C16/C18 (80/20) MES + nonionic surfactant micellar solutions the first precipitate is that of C16-MES. Thus at $C_{A,sat}$, the concentration of free C18-MES monomers, $c_{18}$, should be lower than the solubility limit, $S_{18} = 0.213$ mM. Fig. S4 shows the dependencies of the calculated values of $c_{18}$ at $C_{A,sat}$ for C16/C18 (80/20) MES + Glucopon and C16/C18 (80/20) MES + Pareth-7 on the total surfactant concentration. It is well illustrated that $c_{18} < S_{18}$ for all values of $C_T$.

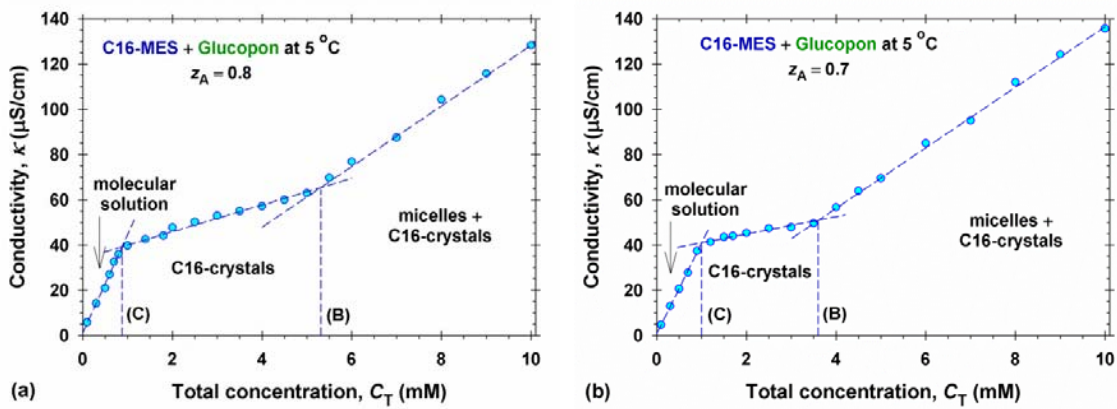

**Fig. S5**. Plots of the electrolytic conductivity of C16-MES + Glucopon solutions versus the total surfactant concentration, $C_T$, at fixed mole fraction of C16-MES: a) $z_A = 0.8$ and b) $z_A = 0.7$. The right and left kinks correspond to intersection of the B and C lines, respectively, of the phase diagram (Fig. 6a).

Fig. S5 shows the measured electrolytic conductivity, $\kappa$, of C16-MES + Glucopon solutions for two fixed mole fractions of C16-MES ($z_A = 0.8$ and $z_A = 0.7$). The ranges of total concentrations, $C_T$, correspond to the two horizontal cross-sections of the phase diagram illustrated in Fig. 6a. The total concentration of the free ions in solutions (Figs. S5-S7) is low (< 10 mM). Therefore: a) the Kohlrausch effect is negligible; b) $\kappa$ is a linear function of the concentration, $C_T$. First, the appearance/disappearance of the mixed micelles (which contain anionic surfactant) at the B-line leads to a kink in conductivity (Fig. S5). In the molecular solution domain, the anionic surfactant is dissociated and the slope of the linear regression corresponds to the molecular conductance of C16-MES at infinite dilution. In the micelle domain, a part of the counterions are bonded to the micelles and the slope of the linear regression decreases. Second, the disappearance of the crystallites of the ionic surfactant (which maintain almost constant the bulk concentration of this component) at the C-line leads to another kink in conductivity (Fig. S5).

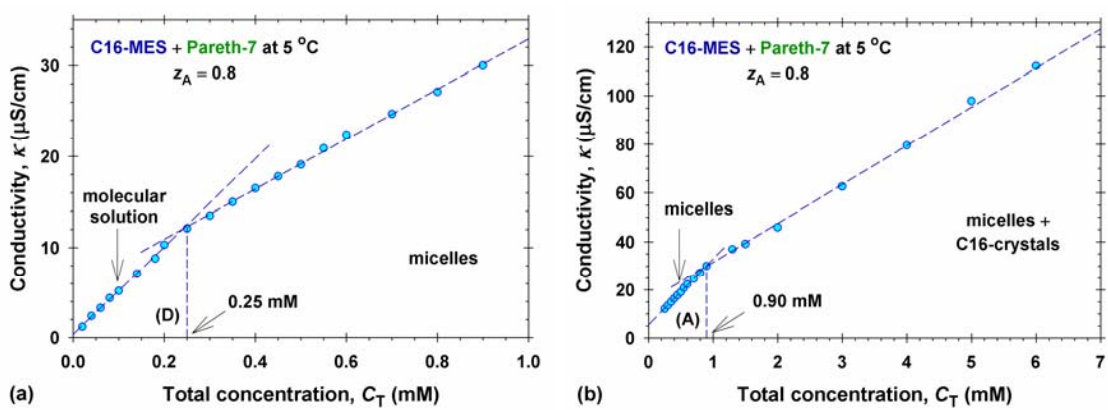

**Fig. S6**. Plots of the electrolytic conductivity of solutions of C16-MES + Pareth-7 versus the total surfactant concentration, $C_T$, at 0.8 mole fraction of C16-MES: a) experimental position of D-line; b) experimental position of A-line in Fig. 6c.

Fig. S6 shows the measured electrolytic conductivity of C16-MES + Pareth-7 micellar solutions at molar ratio, $z_A = 0.8$, for different total surfactant concentrations $C_T$ (see Fig. 6c). With the rise of concentration in the domain of molecular solutions, the electrolytic conductivity, $\kappa$, increases. The appearance of micelles leads to the decrease of the conductivity, because of incorporation of ionic surfactants (C16-MES molecules) in the mixed micelles. The experimental position of D-line is detected as a kink in the experimental dependence $\kappa(C_T)$, see Fig. S6a. With the subsequent increase of $C_T$, one intersects the A-line in Fig. 6c: the phase boundary between the domain with micelles and that with micelles +

C16-crystals. As a result of the precipitation of C16-MES, the slope of the electrolytic conductivity curve decreases. The respective position of the kink point in the experimental dependence $\kappa(C_T)$ shown in Fig. S6b corresponds to the position of A-line in the phase diagram, see Fig. 6c.

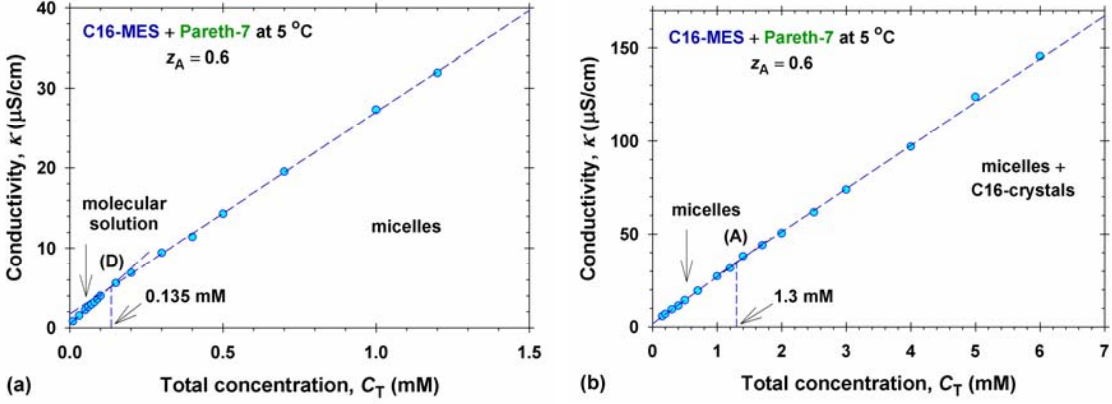

**Fig. S7**. Plots of the electrolytic conductivity of solutions of C16-MES + Pareth-7 versus the total surfactant concentration, $C_T$, at 0.6 mole fraction of C16-MES: a) experimental position of D-line; b) experimental position of A-line in Fig. 6c.

Fig. S7 shows the analogous experiments for electrolytic conductivities of C16-MES + Pareth-7 micellar solutions at lower molar ratio, $z_A = 0.6$. Note that the amount of ionic surfactant is less than those for $z_A = 0.8$ (Fig. 5). For that reason, the kink points in the electrolytic conductivity curve are less pronounced. Nevertheless, they are measurable, but the precisions of obtained values 0.135 mM and 1.3 mM in Figs. S7a and S7b are lower than those given in Fig. S6.

*Rule for ordering of the components A1 and A2*

The molar ratio of the two components in MES mixture, $\lambda$, is defined as follows:

$$\lambda \equiv \frac{C_{A1}}{C_{A2}} = \frac{z_{A1}}{z_{A2}}, \quad z_{A1} = \frac{\lambda z_A}{1+\lambda}, \quad z_{A2} = \frac{z_A}{1+\lambda} \tag{S1}$$

At the D-line, (molecular solution)/micelles, all components are in a molecular form. The total concentration is equal to the CMC of the nonionic surfactant for $z_A = 0$. With the increase of $z_A$, the concentration of one of the components in the MES mixture reaches the solubility limit. Thus the simple rule for the ordering of MES components is the following. The first MES component is that, which satisfies the inequality:

$$\lambda = \frac{C_{A1}}{C_{A2}} > \frac{S_{A1}}{S_{A2}} \tag{S2}$$

For example, the mixture C16/C18 (80/20) MES has $\lambda = 4.301$ and $S_{16}/S_{18} = 3.263$ (Table 1): for this mixture, the first component, A1, is C16-MES and the second one, A2, is C18-MES. For the mixture C16/C18 (60/40) MES, one obtains $\lambda = 1.613$ and $S_{16}/S_{18} = 3.263$: the inequality, Eq. (S2), is not fulfilled. If we change the order of the components, C18/C16 (40/60), then $\lambda = 0.62$, $S_{18}/S_{16} = 0.3065$, and $\lambda > S_{18}/S_{16}$: for this mixture, the first component, A1, is C18-MES and the second – C16-MES.